# A Constraint Programming Model for the Super-agile Earth Observation Satellite Imaging Scheduling Problem

Margarida Caleiras[1][a], Samuel Moniz[1][b] and Paulo Jorge Nascimento[1][c]
[1]*Department of Mechanical Engineering, CEMMPRE, University of Coimbra, Portugal*

Abstract: As the dependence on satellite imaging continues to grow, modern satellites have become increasingly agile, with the new generation, namely super-agile Earth observation satellites (SAEOS), providing unprecedented imaging flexibility. The highly dynamic capabilities of these satellites introduce additional challenges to the scheduling of observation tasks, as existing approaches for conventional agile satellites do not account for variable observation durations and multiple imaging directions. Although some efforts have been made in this regard, the SAEOS imaging scheduling problem (SAEOS-ISP) remains largely unexplored, and no exact approaches have yet been proposed. In this context, this study presents the first exact Constraint Programming formulation for the SAEOS-ISP, considering flexible observation windows, multiple pointing directions and sequence-dependent transition times across multiple satellites. Computational experiments on a newly generated benchmark set demonstrate that the model can be solved efficiently and within very short computational times. Moreover, the results also show that the proposed approach has the potential to achieve higher computational performance compared to the non-exact approaches that are currently considered state-of-the-art.

## 1 INTRODUCTION

In this paper, we address the Earth observation satellite imaging scheduling problem (EOS-ISP), which involves selecting and assigning imaging tasks to satellites in order to maximize mission performance under operational and physical constraints. The scheduling of EOSs has been a growing concern for space agencies and operators for more than three decades, as the number of satellites and customer demands continues to increase faster than the available orbital, energy, and communication capacities (Bensana et al., 1996; Wang et al., 2021; Ferrari et al., 2025). Efficient scheduling is essential to ensure that these limited space resources are effectively managed, especially in time-critical applications such as disaster management, climate monitoring, and security and defense (Chen et al., 2019). However, the growing scale and agility of satellite systems have made scheduling a highly complex and resource-constrained problem that requires fast and reliable decision-making (Ferrari et al., 2025).

Solving the EOS-ISP requires coordinating multiple satellites, each operating under limited target visibility opportunities, energy, and data storage constraints, while ensuring that imaging plans remain feasible and align with mission priorities. These operational dependencies create a complex balance of competing objectives. For instance, maximizing coverage and image quality often conflicts with minimizing mission completion time or maintaining fairness among customers in shared systems. Moreover, unpredictable factors such as weather conditions or last-minute observation requests may require real-time replanning to adapt schedules on short notice. As satellite constellations become larger and more agile, these trade-offs intensify, calling for scheduling approaches that not only handle the intrinsic interdependencies of the EOS-ISP but also produce feasible and high-quality solutions in real time. Building on this, this paper focuses on developing an exact and computationally efficient method for the super-agile EOS-ISP (SAEOS-ISP), where dynamic observation capabilities and time-dependent transition constraints further increase the complexity of real-world operations.

### 1.1 Scheduling of Earth Observation Satellites

Scheduling in Earth observation missions involves deciding when and how satellites acquire images of ground targets while respecting visibility, energy and memory constraints. This problem has been extensively studied in the literature across multiple perspectives and variants, and it has proven to be NP-hard (Lemaıitre et al., 2002).

[a]  https://orcid.org/0009-0008-4815-7265
[b]  https://orcid.org/0000-0002-7813-4514
[c]  https://orcid.org/0000-0003-4139-1100

EOSs can observe targets using onboard instruments such as sensors or cameras, each covering a limited region of the Earth's surface defined by the satellite's ground track, which is the projection of the orbit onto the Earth's surface, and the field of view, that is, the angular width that the sensor can cover. Targets can have different shapes, typically represented as spots (i.e., small circular areas) or polygonal areas, and can be transformed into a set of rectangular strips that cover the entire target (Lemaıitre et al., 2002). The width of each strip corresponds to the width of the satellite's sensor, while its length varies depending on the target's size and geometry (see Figure 1). Each strip thus represents an individual imaging task.

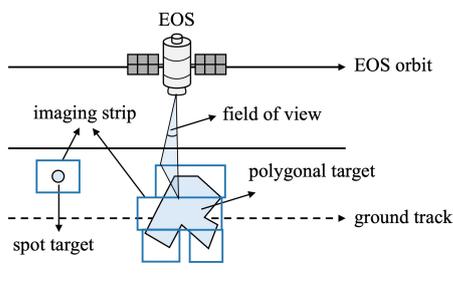

Figure 1: The filed of view, spot target and polygonal target.

A satellite can only observe a strip when it is visible from its orbit, which defines a Visible Time Window (VTW). The actual time interval during which the satellite performs the observation is referred to as the Observation Time Window (OTW). An EOS can complete several orbits around the Earth in a single day, meaning it may have multiple VTW for the same strip in the scheduling horizon (usually 24 hours). For the same strip, multiple VTW from different satellites may overlap. Similarly, for the same satellite, VTW from different strips may overlap. However, each satellite can only perform one imaging task at a time (see Figure 2). In addition to visibility constraints, several operational requirements can be considered, including energy and memory capacity, atmospheric conditions and sensor reorientation. Each observation requires the satellite to adjust its sensor toward the corresponding strip, implying that distinct strips require different orientations. Thus, a transition time between two consecutive scheduled observations has to be considered (Lemaıitre et al., 2002; Liu et al., 2017; Chen et al., 2019).

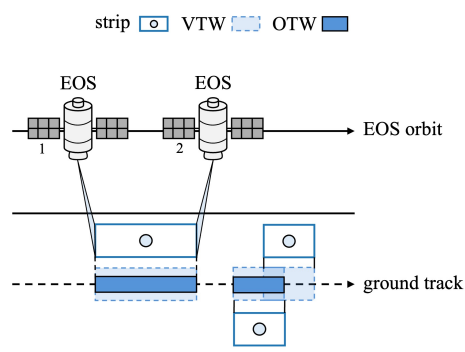

Figure 2: Relation between visible and observation time windows.

Depending on their maneuvering capabilities, satellites can be classified as Conventional EOS (CEOS), Agile EOS (AEOS), or Super-Agile EOS (SAEOS), with AEOS and SAEOS currently being the most used due to their high maneuvering capabilities (Wang et al., 2021). Unlike CEOS, which can only rotate along the roll axis, AEOS have three degrees of freedom (roll, pitch, and yaw), allowing maneuverability during and between image acquisitions (Lemaıitre et al., 2002). As shown in Figure 3, the roll angle defines the rotation of the satellite around its longitudinal axis, enabling lateral pointing toward the strip, whereas the pitch angle controls the satellite's rotation around its transverse axis, allowing the sensor to move upward or downward. The yaw angle describes the satellite's rotation around its vertical axis, adjusting its horizontal orientation over the Earth. In AEOS, the start time of each observation is flexible, and the VTW is typically longer than the actual OTW. For CEOS, the OTW always coincides with its VTW. Although the agile characteristic greatly improves the observation efficiency, the complexity of the scheduling also increases dramatically (Liu et al., 2017; Wang et al., 2021). The difficulty lies essentially in having to decide the start of the observation within a VTW. Moreover, given the higher sensitivity of the pitch angle compared to the roll angle along each VTW (Liu et al., 2017), it becomes necessary to model time-dependent transition times between consecutive tasks, further increasing the complexity of the problem.

SAEOSs are a new generation of agile satellites that has been recently introduced. In contrast to AEOS, which can only acquire strips parallel to the sub-satellite track, SAEOS can perform imaging in arbitrary directions. Furthermore, SAEOS can perform attitude maneuvering (sensor reorientation) while imaging, whereas AEOS can image by orbit motion only (Lu et al., 2022b). This dynamic imaging capabilities relax the need for orientation control during image acquisition, making the imaging duration of each strip variable (subject to angular velocity limits). This high maneuverability expands the optimization search space and increases imaging flexibility, thereby posing new challenges to the scheduling problem and leading to substantially higher computational complexity (Lu et al., 2022b).

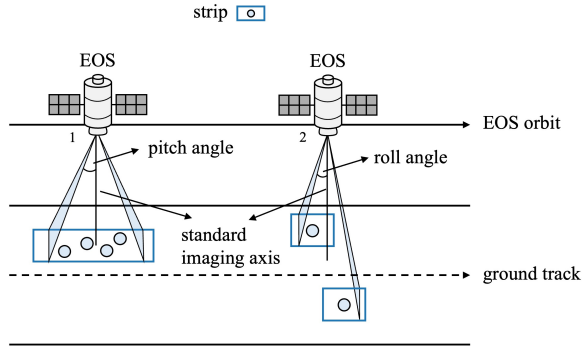

Figure 3: Roll and pitch angle for an EOS.

Most existing studies focus on the AEOS-ISP, for which both exact and non-exact approaches have been developed. However, these methods cannot be directly extended to the SAEOS-ISP due to the additional modeling requirements introduced by variable imaging durations and multiple imaging directions (Lu et al., 2022b). Regarding the SAEOS-ISP, research is still at an early stage, with studies focusing on developing heuristics or meta-heuristics (Ferrari et al., 2025). To the best of our knowledge, no existing method has yet provided optimal solutions or tight upper bounds for this specific problem. Furthermore, the approaches proposed so far report computation times that are not suitable for real-time decision-making, even in small-scale scenarios (Lu et al., 2022a; Lu et al., 2023). Therefore, this study focuses on developing an exact and computationally efficient method for the SAEOS-ISP, aiming to provide optimal or near-optimal solutions within practical computation times.

## 1.2 Methodology

We propose an exact optimization model based on Constraint Programming (CP), capable of solving small to medium-sized instances of the SAEOS-ISP while incorporating often overlooked features such as variable imaging durations, multiple imaging directions, and sequence-dependent transition times. CP is particularly attractive for this problem, as it can produce high-quality solutions within short computational times and has shown remarkable success in efficiently solving complex scheduling problems (Meng et al., 2020; Yunusoglu and Yildiz, 2021; Abreu et al., 2022). Exact optimization approaches have previously been applied to the AEOS-ISP. However, the higher agility of super-agile satellites, particularly their ability to perform active imaging and acquire observations in multiple directions, makes the SAEOS-ISP more complex and challenging to solve. The main differences lies in the fact that imaging durations for the same strip may vary, and the imaging direction itself becomes a decision variable. Nevertheless, both problems share several core features, such as visibility and task constraints (Lemaıitre et al., 2002; Lu et al., 2023) and transition times (Liu et al., 2017).

Compared to previous exact methods developed for the AEOS-ISP, our model accounts for: (1) variable-length observation windows, allowing imaging durations to vary for the same strip; (2) multiple imaging directions, making the pointing direction a decision variable; and (3) sequence-dependent transition times, which capture satellite maneuvers between consecutive observations. Although specifically designed for the SAEOS-ISP, the proposed model can also be directly applied to the AEOS-ISP. To assess its performance, we evaluated the approach on a newly generated benchmark dataset composed of multiple multi-satellite scenarios and multi-type targets.

## 1.3 Contributions

The main contributions of this study are as follows. Firstly, to the best of our knowledge, this is the first exact CP formulation proposed for the SAEOS-ISP. The model can be directly applied to both agile and super-agile imaging scheduling scenarios, supporting different target types and mission configurations. Secondly, we introduce a new benchmark dataset and conduct a comprehensive computational study to evaluate the performance of the proposed approach. Finally, we present the first comparative analysis with other methods for this problem, namely the non-exact approach proposed in (Lu et al., 2023).

The remainder of this paper is organized as follows. Section 2 reviews the related literature on agile and super-agile satellite imaging scheduling. Section 3 describes the problem and its main characteristics. The proposed constraint programming formulation is presented in Section 4, followed by the experimental setup and computational results in Section 5. Finally, Section 6 concludes the paper and outlines directions for future research.

## 2 LITERATURE REVIEW

Most studies on the AEOS-ISP focus on heuristic or metaheuristic approaches, although exact methods have also been proposed. Among the exact methods, common approaches include dynamic programming (Lemaître et al., 2002; Peng et al., 2020), branch-and-bound (Chu et al., 2017), and branch-and-price algorithms (Li et al., 2024; Peng et al., 2025). Other works adopt mathematical programming approaches, including Constraint Programming (Lemaître et al., 2002) and Mixed-Integer (Linear) Programming formulations (Chen et al., 2019; Wang et al., 2023; Berger et al., 2020), which are solved using commercial solvers. As for non-exact approaches, a wide variety of heuristics and metaheuristics have been proposed, ranging from greedy and local search methods (Lemaître et al., 2002; Tangpattanakul et al., 2015) and tabu search (Cordeau and Laporte, 2005; Habet et al., 2010), to constructive methods and ant colony optimization (Xu et al., 2016). More recent trends include adaptive large neighborhood search (Liu et al., 2017; He et al., 2018), evolutionary and genetic algorithms (Li et al., 2018; Niu et al., 2018; Xu et al., 2018; Zhibo et al., 2021), as well as temporal conflict network-based heuristics (Xie et al., 2019) and greedy randomized iterated local search (Peng et al., 2022). Some of these works address different variants of the AEOS-ISP, such as single-orbit scheduling (Lemaître et al., 2002; Cordeau and Laporte, 2005) or multi-objective formulations (Habet et al., 2010; Tangpattanakul et al., 2015). Others focus on additional problem attributes, including time-dependent transition times (Liu et al., 2017; He et al., 2018; Peng et al., 2022), time-dependent profits (Peng et al., 2020), and interval-varying profits (Li et al., 2024). Overall, these studies provide valuable modeling insights and solution approaches for understanding how to tackle the EOS-ISP. However, they are specifically designed for the AEOS-ISP. Despite some conceptual similarities with the SAEOS-ISP, particularly in terms of visibility modeling and scheduling objectives, these methods cannot be directly applied to this problem, as they do not account for imaging direction variables or active imaging capabilities, that is, both the imaging direction and the imaging duration are assumed to be fixed.

Regarding the SAEOS-ISP, only a few relevant studies have been reported in the literature. In the single-satellite context, (Yang et al., 2018) proposed a bi-objective optimization model and solved it using a hybrid multi-objective evolutionary algorithm. (Cui et al., 2018) considered a single-objective formulation with a planning horizon shorter than one orbit period, and solved it using an ant colony optimization algorithm with tabu lists. (Chang et al., 2020) were the first to consider the imaging duration of the strip as a variable of the problem. The authors proposed a dedicated method to calculate the minimum image duration of each observation, considering task priority and ground target congestion, and solved the model using a greedy heuristic. (Lu et al., 2022b; Lu et al., 2022a; Lu et al., 2023; Lu et al., 2024) also considered a variable imaging duration, employing an improved particle swarm optimization algorithm to solve the single-orbit problem. (Lu et al., 2022b) addressed the single-satellite case, while (Lu et al., 2022a) introduced a new strip decomposition method for various target types considering dynamic imaging capabilities. (Lu et al., 2023) extended the formulation to the multi-satellite scenario, and (Lu et al., 2024) investigated the impact of dynamic imaging on the imaging quality of SAEOS. Finally, (Wu et al., 2024) proposed the first model considering energy and memory capacity constraints, although imaging direction and variable durations were not modeled. The authors implemented an improved adaptive large neighborhood search algorithm based on a two-stage framework to solve the problem.

To the best of our knowledge, no exact method has yet been developed to deal with the SAEOS-ISP. While the non-exact approaches proposed in the literature have demonstrated good performance in finding satisfactory solutions, they generally require extensive parameter tuning and cannot guarantee solution quality or optimality. Moreover, the few studies that report computational performance indicate run times on the order of $10^3$ seconds even for relatively small instances (Lu et al., 2022a; Lu et al., 2023), which suggests that they may become computationally inefficient for more complex scenarios. In view of these challenges, this paper proposes a multi-satellite constraint programming formulation for the SAEOS-ISP, considering variable imaging durations and sequence-dependent transition times between consecutive observations. The proposed approach enables efficient allocation of multi-satellite resources and the simultaneous determination of the observation sequence for each satellite.

## 3 THE EOS IMAGING SCHEDULING PROBLEM

We consider the scheduling of multiple imaging tasks assigned to a constellation of satellites, taking into account visibility and task constraints, and sequence-dependent transition times.

The satellites $i \in I = \{1,\ldots,n_I\}$ are operated by a ground station that collects imaging requests from customers. Each request involves the observation of a target $r \in R = \{1,\ldots,n_R\}$. Due to image-processing and satellite observation constraints, targets are decomposed into strips $j \in J = \{1,\ldots,n_J\}$. After decomposition, a unique strip scheme is obtained for each target with a fixed direction, and therefore each strip can be imaged in one of two possible imaging senses, the orbital sense ($l=0$) or the opposite-orbital sense ($l=1$) (see Figure 4). Each satellite $i$ completes multiple orbits $k \in K_i$ within the scheduling horizon, thereby generating a set of VTW $VW_{ij}$ with respect to each strip $j$. We denote by $vw_{ijkl}$ the VTW of satellite $i$ for strip $j$ on orbit $k$ with imaging sense $l$, which is composed of a tuple $vw_{ijkl} = \{vws_{ijkl}, vwe_{ijkl}\}$, where $vws_{ijkl}$, $vwe_{ijkl}$ denote the start time and the end time of $vw_{ijkl}$, respectively. Each strip $j$, characterized by a weight $w_j$, must be observed continuously for a processing time $p_{ijkl}$ within one of the available VTW. The weights of the strips are derived from the weights assigned to their corresponding targets, which are defined by the customer and represent observation priorities, with higher values indicating higher importance. Further, in each VTW $vw_{ijkl}$, strip $j$ is characterized by two endpoints, $sp^a_{ijkl}$ and $sp^b_{ijkl}$. A strip is observed if both endpoints are observed in the same $vw_{ijkl}$ (see Figure 4). Further, each observation is defined by the roll, pitch, and yaw angles adopted by the satellite during imaging. Therefore, for each satellite $i$ we must account for the attitude maneuvering times $\Delta_{(ijkl),(ij'k'l')}$, referred to as transition times, between two consecutively observed strips $j$ and $j'$ in their respective VTW $vw_{ijkl}$ and $vw_{ij'k'l'}$.

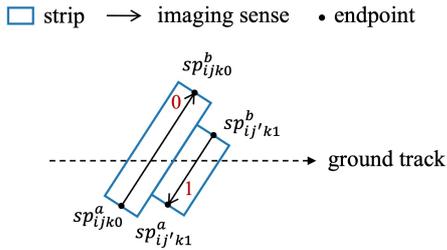

Figure 4: Imaging sense and endpoints of strips.

The SAEOS-ISP considered in this paper consists of determining, for each satellite, which strips to observe, in which orbit, and in which imaging sense, while satisfying visibility constraints, and transition times. Based on the practical applications in reality, we made the following definitions and assumptions:

1. the satellites mentioned in this paper are SAEOSs that can perform imaging with simultaneous attitude maneuvering;

2. satellite orbits are deterministic and known in advance, so the VTW of strips can be computed a priori;

3. it is assumed that the elevation angle of the Sun always satisfies the requirements of optical imaging;

4. atmospheric effects and weather interference are not considered in the model;

5. targets are decomposed into strips. Each strip must be fully acquired in a single observation; otherwise, the observation of the strip is considered a failure;

6. each strip can only be imaged once, by a single satellite, and its maximum expected observation count is one;

7. no new insertion requests are considered during the imaging process of the satellites;

8. the download of the acquired images is not considered in this study;

9. the on-board memory and energy of the satellites are assumed to be sufficient; thus, the data storage and energy consumption are not considered in the proposed model;

10. the scheduling problem is oversubscribed and not all customer requests can be satisfied. Thus it is acceptable that some strips are not assigned.

# 4 THE CONSTRAINT PROGRAMMING MODEL

The proposed CP model was developed using IBM ILOG CP Optimizer, which provides a comprehensive set of built-in functions and global constraints specifically designed for scheduling applications. The notation adopted in the following section follows the CP Optimizer modeling syntax (IBM Corporation, 2024).

## 4.1 Notation

This subsection describes and explains the sets, parameters, and variables that compose the CP model.

The VTW are obtained considering the characteristics of the satellites' orbits and their maneuvering capabilities, namely the field of view and the maximum maneuvering angle. The set of strips is obtained through target decomposition, which is itself a highly complex combinatorial optimization problem, usually referred to as the strip covering problem

(Hu et al., 2021). For polygonal targets, we adopted the decomposition method proposed by (Lu et al., 2022a), which, for each target, determines the orientation angle of the set of strips that minimizes the total strip length. For spot targets, we considered a single strip per target with a random orientation. Although target decomposition is a problem closely related to our study, its resolution lies beyond the scope of this work, and therefore for further details on the problem definition and the adopted method, please refer to (Hu et al., 2021; Lu et al., 2022a).

In our model, processing times are treated as variable. However, each strip's processing time is bounded below by the satellite's maximum angular velocity, that is, the minimum time required to observe the strip, and above by the duration of the corresponding VTW. Hence, we have

$$p_{ijkl}^U = vwe_{ijkl}^U - vws_{ijkl}^L, \qquad (1)$$

and, considering the continuous attitude adjustment mode (Lu et al., 2022b),

$$p_{ijkl}^L = \delta_i + \max\left\{\frac{\Delta\varphi_{ijkl}^{1,2}}{\bar{\varphi}_i}, \frac{\Delta\omega_{ijkl}^{1,2}}{\bar{\omega}_i}, \frac{\Delta\kappa_{ijkl}^{1,2}}{\bar{\kappa}_i}\right\}. \qquad (2)$$

Here $\delta_i$ denotes the settling time of satellite $i$ (the time required to turn the satellite sensor on and off), and $\bar{\varphi}_i, \bar{\omega}_i$ and $\bar{\kappa}_i$ are the average angular velocities of the satellite along the corresponding axes, respectively. $\Delta\varphi_{ijkl}^{1,2}, \Delta\omega_{ijkl}^{1,2}, \Delta\kappa_{ijkl}^{1,2}$ are the angular displacements during attitude maneuvering from strip endpoint $sp_{ijkl}^1$ to $sp_{ijkl}^2$ along the roll, pitch, and yaw axes, where $sp_{ijkl}^1$ and $sp_{ijkl}^2$ denote the first and second endpoint to be observed in VTW $vw_{ijkl}$, respectively. Note that in VTW $vw_{ijk0}$, $sp_{ijk0}^1 = sp_{ijk0}^a$ and $sp_{ijk0}^2 = sp_{ijk0}^b$, whereas in VTW $vw_{ijk1}$, $sp_{ijk1}^1 = sp_{ijk1}^b$ and $sp_{ijk1}^2 = sp_{ijk1}^a$. In a similar way, we define the transition times as follows

$$\Delta_{(ijkl),(ij'k'l')} = \delta_i + \max\left\{\frac{\Delta\varphi_{(ijkl),(ij'k'l')}^{2,1}}{\bar{\varphi}_i}, \frac{\Delta\omega_{(ijkl),(ij'k'l')}^{2,1}}{\bar{\omega}_i}, \frac{\Delta\kappa_{(ijkl),(ij'k'l')}^{2,1}}{\bar{\kappa}_i}\right\}. \qquad (3)$$

where $\Delta\varphi_{(ijkl),(ij'k'l')}^{2,1}, \Delta\omega_{(ijkl),(ij'k'l')}^{2,1}$ and $\Delta\kappa_{(ijkl),(ij'k'l')}^{2,1}$ are the angular displacements during attitude maneuvering from strip endpoint $sp_{ijkl}^2$ to $sp_{ij'k'l'}^1$ along the roll, pitch, and yaw axes, respectively.

Interval variables are intrinsic variables of the CP Optimizer dedicated to scheduling, and represent an interval of time during which an activity occurs. These variables are defined by a start time, a processing time, and an end time, where the start time plus the processing time equals the end time. Thus, in the case of $X_{ijkl}$, it provides the start, processing time (which in our case is variable), and end time of strip $j$. A key feature is that interval variables can be optional, which is why $X_{ijkl}$ only exists if strip $j$ is observed by satellite $i$ in VTW $vw_{ijkl}$, and $Y_j$ only exists if there is at least one observation for strip $j$.

Sequence variables are also inherent variables of the CP Optimizer dedicated to scheduling, and their function is to determine the order of a set of interval variables, which in this case are the variables $X_{ijkl}$ of each satellite $i$. Intrinsically, variables $X_{ijkl}$ that take value zero are not to be considered in the ordering.

| Sets | |
|---|---|
| $i \in I$ | set of satellites, $I = \{1, \ldots, n_I\}$ |
| $r \in R$ | set of targets, $R = \{1, \ldots, n_R\}$ ($R = R_s \cup R_p$, where $R_s$ and $R_p$ denote the sets of spot and polygonal targets, respectively) |
| $j \in J$ | set of strips, $J = \{1, \ldots, n_J\}$ ($J = J_s \cup J_p$, where $J_s$ and $J_p$ denote the sets of strips derived from spot and polygonal targets, respectively) |
| $k \in K_i$ | set of orbits of satellite $i$, $K_i = \{1, \ldots, n_{K_i}\}$ |
| $l \in L$ | set of imaging directions, $L = \{0, 1\}$ |
| $VW_{ij}$ | set of VTW of satellite $i$ for strip $j$, $VW_{ij} = \{vw_{ijkl} | k \in K_i, l \in L\}$ ($VW_{ij} = \emptyset$ if satellite $i$ has no VTW for strip $j$) |
| **Parameters** | |
| $w_r, w_j$ | weights of target $r$ and strip $j$, respectively |
| $A_r, A_j$ | area of target $r$ and strip $j$, respectively |
| $p_{ijkl}^L, p_{ijkl}^U$ | lower and upper bound, respectively, for the processing time of strip $j$ in VTW $vw_{ijkl}$ |
| $\Delta_{(ijkl),(ij'k'l')}$ | minimal transition time required for switching from strip $j$ (observed in VTW $vw_{ijkl}$) to strip $j'$ (observed in VTW $vw_{ij'k'l'}$) on satellite $i$ |
| $vws_{ijkl}^L, vws_{ijkl}^U$ | lower and upper bound, respectively, for the start time of VTW $vw_{ijkl}$ |
| $vwe_{ijkl}^L, vwe_{ijkl}^U$ | lower and upper bound, respectively, for the end time of VTW $vw_{ijkl}$ |
| **Variables** | |
| $X_{ijkl}$ | optional interval variable denoting the time interval during which strip $j$ is observed by satellite $i$ in VTW $vw_{ijkl}$. The domain of $X_{ijkl}$ is defined by $\text{start}(X_{ijkl}) \in [vws_{ijkl}^L, vws_{ijkl}^U]$, $\text{end}(X_{ijkl}) \in [vwe_{ijkl}^L, vwe_{ijkl}^U]$, $\text{size}(X_{ijkl}) \in [p_{ijkl}^L, p_{ijkl}^U]$, and $\text{end}(X_{ijkl}) = \text{start}(X_{ijkl}) + \text{size}(X_{ijkl})$. |
| $Y_j$ | optional interval variable denoting the interval of time during which strip $j$ is observed. |

$S_i$    sequence variable describing the order of the strips assigned to satellite $i$, that is, $S_i$ on $[X_{ijkl}]_{j \in J, k \in K_i, l \in L}, \forall i \in I$

## 4.2 Model Formulation

The CP formulation can be stated as follows:

$$\text{Maximize} \sum_{j \in J_s} w_j \times presenceOf(Y_j) + \sum_{r \in R_p} w_r \times f\left(\sum_{j \in J_p} \frac{1}{A_r} A_j \times presenceOf(Y_j)\right) \quad (4)$$

Subject to

$$alternative\left(Y_j, \{X_{ijkl} \mid i \in I, k \in K_i, l \in L\}\right), \forall j \in J \quad (5)$$

$$noOverlap\left(S_i, \Delta_{(ijkl),(ij'k'l')}\right), \forall i \in I \quad (6)$$

Objective function (4) aims to maximize the total observation coverage revenue. The function is composed of two parts, the summed profits for observed spot targets and the summed profits for polygonal targets. The first component is calculated as the sum of the weights of the strips corresponding to spot targets that are successfully scheduled for observation. This is achieved by using the built-in function $presenceOf(Y_j)$ of the CP Optimizer, which returns 1 if strip $j$ is observed and 0 otherwise. The second component accounts for polygonal targets, where the profit of each scheduled strip is proportional to the percentage of the target's area covered by that strip (Wang et al., 2011). The total profit for a partially observed polygon is modeled as a piecewise linear function $f$ of the cumulative covered surface ratio $x \in [0, 1]$, where $x$ represents the proportion of the polygon's total area that has been observed. The function $f$ is defined by

$$f(x) = \begin{cases} 0.25x, & 0 \leq x < 0.4, \\ 1.0x - 0.3, & 0.4 \leq x < 0.7, \\ 2.0x - 1.0, & 0.7 \leq x \leq 1. \end{cases} \quad (7)$$

Unlike the linear formulation used for spot targets, this representation, originally proposed as a modeling approach by (Lemaître et al., 2002), penalizes partial observations while strongly encouraging complete coverage of each polygon. In our implementation, $f$ is directly expressed in the CP Optimizer using its built-in operator *piecewiseLinear*.

The scheduling constraints make use of the CP Optimizer's global constraints *alternative*() and *noOverlap*(). In constraints (5) a satellite $i$ from the set of available satellites is scheduled to observe strip $j$ in VTW $vw_{ijkl}$. Then, constraints (6) are responsible for defining the sequence of the different strips $j$ assigned to satellite $i$, guaranteeing that only one strip is observed at a time and embedding setup times, $\Delta_{(ijkl),(ij'k'l')}$, between each pair of consecutive observations.

## 5 EXPERIMENTAL STUDY

In the following, we describe our experiments on several problem instances. The computational results were obtained using a MacBook Pro 14" with an Apple M3 processor (8 cores) and 16GB of RAM. The CP model was solved using IBM ILOG CP Optimizer 22.1.1 with the Python 3.11.13 API DOcplex.cp library version 2.30.251.

### 5.1 Method for Generating Instances

To evaluate the performance of our model, we generated several test instances. As no standard datasets are available for super-agile satellite scheduling, we adapted the benchmark generation methods proposed by (Wang et al., 2011; Liu et al., 2017; Chen et al., 2019), originally designed for agile satellites.

The experimental analysis considers a four-satellite constellation, with the corresponding parameters presented in Table 1. These parameters specify the satellites' orbital configuration, including the orbit semi-major axis, eccentricity, inclination, right ascension of the ascending node (RAAN), argument of perigee, and mean anomaly, as well as their attitude and imaging characteristics, such as the maximum maneuvering angle, angular velocity, field of view, and settling time. Each satellite has the same parameters, except for the mean anomaly. The maximum maneuvering angle is assumed to be the same across all three axes. The start time of the scheduling period is 01-09-2025 00:00:00, and the considered scheduling horizon is 24 hours.

Table 1: Parameters of each satellite.

| Parameter | Sat1 | Sat2 | Sat3 | Sat4 |
|---|---|---|---|---|
| Orbit semi-major axis (km) | | 6998 | | |
| Eccentricity | | 0.001 | | |
| Inclination (°) | | 97.9 | | |
| RAAN (°) | | 0.0 | | |
| Argument of perigee (°) | | 0.0 | | |
| Mean anomaly (°) | 90 | 95 | 100 | 105 |
| Maximum maneuvering angle (°) | | 60 | | |
| Angular velocity (°/s) | | 6 | | |
| Field of view (°) | | 1.3 | | |
| Settling time (s) | | 4 | | |

To generate scenarios with potential conflicts among VTW, the observation targets were randomly generated within a narrow region, bounded by latitudes 75°S–65°S and longitudes 75°W–65°W. We considered seven different instance configurations (A-G), varying in the number and type of targets. The generation procedure can be summarized as follows. Each spot target corresponds to a single point defined by latitude and longitude coordinates, while each polygonal target is represented as an irregular polygon with five to six vertices. The coordinates of both the spot targets and the polygon centers were randomly generated within the defined region. The weights of the spot targets were randomly sampled as integers in the range [1, 10], whereas the weight of each polygonal target (with area denoted by $A$) was set to $(A/25)$ multiplied by an integer randomly selected from [1, 10] (Wang et al., 2011). The VTW of these targets, as well as the corresponding observation angles and transition times, were obtained through visibility analysis.

For each configuration of targets, ten different instances were generated. Table 2 presents the details of these instances.

## 5.2 Computational Results

The experimental study conducted in this section has two main goals. The first is to provide a preliminary assessment of the performance of our CP model. To this end, small to medium-sized instances are used, and the results are analyzed in terms of computation time and solution quality. The second objective is to conduct a comparison with existing approaches, in order to assess the ability of our model to achieve computational performance comparable to state-of-the-art non-exact methods. We note that our comparative analysis was conducted only with reference to the study presented in (Lu et al., 2023). As discussed in Section 2, there are still relatively few works addressing the SAEOS-ISP, and those that exist differ from our formulation in several key aspects. In particular, many studies neglect either the variability in imaging duration or in imaging direction (Yang et al., 2018; Cui et al., 2018; Wu et al., 2024). Among the existing works, (Lu et al., 2022b; Lu et al., 2022a; Lu et al., 2023; Lu et al., 2024) are the most comprehensive in this regard. The models proposed in (Lu et al., 2022b; Lu et al., 2022a; Lu et al., 2023) are essentially identical, differing mainly in the number of satellites and the number and type of targets considered, while (Lu et al., 2024) introduces dynamic imaging capabilities that fall outside the scope of our study. Therefore, we present a comparison with (Lu et al., 2023), which proposes a multi-satellite model for super-agile satellites tasked with observing a large polygonal target within a single orbit.

### 5.2.1 Optimization Results

In this section we analyze the main results of the model. A computation time limit (TL) of 1 hour was set. The results are shown in Table 2. Here, the total profit corresponds to the sum of the profits obtained for all scheduled observations, expressed as a percentage of the total possible profit according to the linear version of the objective function; in other words, it measures how much of the potential profit was effectively achieved by the schedule, regardless of the type of target. The makespan represents the time of the last scheduled observation, indicating the overall duration of the schedule, the optimality gap shows the relative difference between the best feasible solution and the best bound found by the solver, and the run time corresponds to the total computational time required to obtain the reported solution. Note that a gap value of 0% indicates that the solution is optimal and that the solver has proven its optimality. For all other cases, the reported solutions are feasible.

As shown in Table 2, the model is able to find optimal solutions within short computational times for instances of types A–E. For the larger instance types F and G, the solver was still able to produce feasible solutions, although optimality could not always be proven within the imposed time limit. As expected, the run time increases with the instance size, which can be explained by the structural growth of the model itself. As shown in Table 4, increasing the number of targets and strips leads to a rapid expansion in the number of decision variables and constraints. Each additional strip introduces new optional intervals and sequencing relations, which significantly enlarge the search space explored by the solver. As a result, larger instances demand higher computational effort and memory, making it increasingly difficult to guarantee optimality within reasonable time limits. Furthermore, it is important to note that even optimal solutions do not guarantee that all strips are scheduled within the considered horizon. This may result from the absence of VTW for certain strips or from the high level of conflict between VTW.

Table 4: Average number of interval variables and constraints per instance group.

| Ins. | Intr. Var. | Const. |
|---|---|---|
| A | 353 | 29 |
| B | 736 | 54 |
| C | 1074 | 80 |
| D | 934 | 81 |
| E | 1865 | 134 |
| F | 1688 | 141 |
| G | 2329 | 191 |

The obtained results highlight the efficiency and robustness of the proposed model for small to medium-sized instances, while also revealing the scalability challenges inherent to the SAEOS-ISP. It is worth emphasizing, however, that for most of the tested instances, the CP model was able to prove optimality within very short computational times. Even in cases where optimality could not be proven, the solver still produced high-quality feasible solutions, with optimality gaps below 5%, for instances of type F, and 10%, for instances of type G.

### 5.2.2 Comparative Analysis

In this section, we present a comparative analysis with the results reported in (Lu et al., 2023). To simulate a scenario similar to that described in the study, we consider a single large polygonal target to be observed by the four-satellite constellation introduced in Section 5.1, within a single orbit. For these experiments, we ensured that in the selected orbit all satellites have valid VTW for all strips of the target, thus replicating the coverage conditions assumed in (Lu et al., 2023).

It is important to note that (Lu et al., 2023) adopted a multi-objective formulation that aims to maximize the total observation coverage revenue while simultaneously minimizing the makespan. To enable a fair comparison, we implemented a similar objective structure in our CP model. Specifically, we defined a lexicographic objective function using CP Optimizer *minimize_static_lex()* built-in-function, introduced in Equation (8), where the primary component maximizes the total observation profit and the secondary component minimizes the makespan $C_{\max}$, defined in Equation (9) as the latest completion time among all satellites. This formulation prioritizes maximizing observation coverage first and then minimizing the overall mission duration. The other constraints are identical to those in the base model.

Table 2: Experimental results

| | Objective function: maximize total observation coverage revenue | | | | | | | | | | | | |
|---|---|---|---|---|---|---|---|---|---|---|---|---|---|
| Instance analysis | | | | | | CP results | | | | | | | |
| Ins. | CtSpot | CtPoly | CtStr | CtVtw | CtStrVtw | CtAcSat | CtAcOrb | CtObSpot | CtObPoly | CtObStr | TotProf(%) | Makespan | Gap(%) | Time(s) |
| A-1 | - | 1 | 22 | 352 | 22 | 3/4 | 1/2 | - | 1/1 | 22/22 | 100.0 | 07:21:40 | 0.0 | 0.02 |
| A-2 | - | 1 | 23 | 368 | 23 | 4/4 | 1/2 | - | 1/1 | 23/23 | 100.0 | 07:22:54 | 0.0 | 0.01 |
| A-3 | - | 1 | 20 | 320 | 20 | 2/4 | 1/2 | - | 1/1 | 20/20 | 100.0 | 07:20:58 | 0.0 | 0.01 |
| A-4 | - | 1 | 21 | 336 | 21 | 4/4 | 1/2 | - | 1/1 | 21/21 | 100.0 | 07:22:34 | 0.0 | 0.01 |
| A-5 | - | 1 | 19 | 304 | 19 | 2/4 | 1/2 | - | 1/1 | 19/19 | 100.0 | 07:20:17 | 0.0 | 0.01 |
| A-6 | - | 1 | 28 | 436 | 28 | 4/4 | 1/2 | - | 1/1 | 28/28 | 100.0 | 07:22:18 | 0.0 | 0.02 |
| A-7 | - | 1 | 17 | 138 | 17 | 3/4 | 1/2 | - | 1/1 | 17/17 | 100.0 | 07:19:46 | 0.0 | 0.01 |
| A-8 | - | 1 | 36 | 416 | 36 | 4/4 | 2/2 | - | 1/1 | 36/36 | 100.0 | 16:52:17 | 0.0 | 0.02 |
| A-9 | - | 1 | 33 | 336 | 33 | 3/4 | 1/3 | - | 1/1 | 33/33 | 100.0 | 07:20:49 | 0.0 | 0.01 |
| A-10 | - | 1 | 34 | 272 | 34 | 4/4 | 1/1 | - | 1/1 | 34/34 | 100.0 | 07:22:11 | 0.0 | 0.04 |
| B-1 | 50 | - | 50 | 664 | 50 | 3/4 | 3/4 | - | 50/50 | 50/50 | 100.0 | 16:50:54 | 0.0 | 0.05 |
| B-2 | 50 | - | 50 | 694 | 50 | 3/4 | 3/4 | - | 50/50 | 50/50 | 100.0 | 16:48:35 | 0.0 | 0.04 |
| B-3 | 50 | - | 50 | 704 | 50 | 3/4 | 3/4 | - | 50/50 | 50/50 | 100.0 | 16:48:47 | 0.0 | 0.06 |
| B-4 | 50 | - | 50 | 638 | 49 | 3/4 | 3/4 | - | 49/50 | 49/50 | 96.8 | 16:51:05 | 0.0 | 0.05 |
| B-5 | 50 | - | 50 | 712 | 50 | 3/4 | 3/4 | - | 50/50 | 50/50 | 100.0 | 16:51:18 | 0.0 | 0.04 |
| B-6 | 50 | - | 50 | 712 | 50 | 3/4 | 3/4 | - | 50/50 | 50/50 | 100.0 | 16:50:33 | 0.0 | 0.05 |
| B-7 | 50 | - | 50 | 672 | 50 | 4/4 | 2/3 | - | 50/50 | 50/50 | 100.0 | 16:50:57 | 0.0 | 0.04 |
| B-8 | 50 | - | 50 | 682 | 50 | 4/4 | 2/3 | - | 50/50 | 50/50 | 100.0 | 16:51:00 | 0.0 | 0.04 |
| B-9 | 50 | - | 50 | 686 | 50 | 4/4 | 2/3 | - | 50/50 | 50/50 | 100.0 | 16:50:56 | 0.0 | 0.04 |
| B-10 | 50 | - | 50 | 698 | 50 | 3/4 | 3/4 | - | 50/50 | 50/50 | 100.0 | 16:51:11 | 0.0 | 0.05 |
| C-1 | 50 | 1 | 70 | 945 | 70 | 4/4 | 3/4 | 50/50 | 1/1 | 70/70 | 100.0 | 16:51:25 | 0.0 | 0.19 |
| C-2 | 50 | 1 | 73 | 1042 | 73 | 4/4 | 3/4 | 50/50 | 1/1 | 73/73 | 100.0 | 16:51:31 | 0.0 | 0.22 |
| C-3 | 50 | 1 | 73 | 927 | 73 | 4/4 | 3/4 | 50/50 | 1/1 | 73/73 | 100.0 | 16:52:53 | 0.0 | 0.18 |
| C-4 | 50 | 1 | 70 | 930 | 70 | 4/4 | 2/3 | 50/50 | 1/1 | 70/70 | 100.0 | 15:18:53 | 0.0 | 0.17 |
| C-5 | 50 | 1 | 95 | 1128 | 95 | 4/4 | 4/4 | 50/50 | 1/1 | 95/95 | 100.0 | 16:51:34 | 0.0 | 1.59 |
| C-6 | 50 | 1 | 76 | 795 | 61 | 4/4 | 4/4 | 50/50 | 1/1 | 61/76 | 36.4 | 16:55:34 | 0.0 | 0.16 |
| C-7 | 50 | 1 | 77 | 1024 | 77 | 4/4 | 3/4 | 50/50 | 1/1 | 77/77 | 100.0 | 16:53:50 | 0.0 | 0.22 |
| C-8 | 50 | 1 | 76 | 1099 | 76 | 4/4 | 3/3 | 50/50 | 1/1 | 76/76 | 100.0 | 16:53:20 | 0.0 | 0.22 |
| C-9 | 50 | 1 | 77 | 1003 | 77 | 4/4 | 3/4 | 50/50 | 1/1 | 77/77 | 100.0 | 16:51:00 | 0.0 | 0.23 |
| C-10 | 50 | 1 | 74 | 1084 | 74 | 4/4 | 3/3 | 50/50 | 1/1 | 74/74 | 100.0 | 16:51:05 | 0.0 | 0.08 |
| D-1 | - | 3 | 76 | 816 | 75 | 4/4 | 4/5 | - | 3/3 | 75/76 | 99.0 | 16:54:58 | 0.0 | 0.16 |
| D-2 | - | 3 | 96 | 842 | 74 | 4/4 | 2/2 | - | 3/3 | 74/96 | 76.5 | 16:55:01 | 0.0 | 0.13 |
| D-3 | - | 3 | 71 | 816 | 71 | 4/4 | 3/3 | - | 3/3 | 71/71 | 100.0 | 16:53:15 | 0.0 | 7.11 |
| D-4 | - | 3 | 70 | 748 | 70 | 4/4 | 2/3 | - | 3/3 | 70/70 | 100.0 | 16:51:13 | 0.0 | 0.29 |
| D-5 | - | 3 | 66 | 615 | 54 | 4/4 | 2/3 | - | 3/3 | 54/66 | 90.5 | 16:50:13 | 0.0 | 0.11 |
| D-6 | - | 3 | 73 | 985 | 73 | 4/4 | 2/3 | - | 3/3 | 73/73 | 100.0 | 16:52:36 | 0.0 | 1.00 |
| D-7 | - | 3 | 82 | 978 | 82 | 4/4 | 2/2 | - | 3/3 | 82/82 | 100.0 | 16:54:04 | 0.0 | 0.08 |
| D-8 | - | 3 | 74 | 1048 | 74 | 4/4 | 2/2 | - | 3/3 | 74/74 | 100.0 | 16:51:54 | 0.0 | 0.09 |
| D-9 | - | 3 | 72 | 712 | 62 | 4/4 | 2/3 | - | 3/3 | 62/72 | 79.1 | 16:51:15 | 0.0 | 0.51 |
| D-10 | - | 3 | 93 | 1006 | 86 | 4/4 | 3/4 | - | 3/3 | 86/93 | 92.0 | 16:54:04 | 0.0 | 0.22 |
| E-1 | 50 | 3 | 147 | 1913 | 146 | 4/4 | 4/4 | 50/50 | 3/3 | 146/147 | 99.8 | 16:54:08 | 0.0 | 3.14 |
| E-2 | 50 | 3 | 127 | 1686 | 126 | 4/4 | 4/4 | 50/50 | 3/3 | 126/127 | 98.3 | 16:53:47 | 0.0 | 9.84 |
| E-3 | 50 | 3 | 131 | 1758 | 131 | 4/4 | 3/4 | 50/50 | 3/3 | 131/131 | 100.0 | 16:53:52 | 0.0 | 2.59 |
| E-4 | 50 | 3 | 118 | 1597 | 118 | 4/4 | 2/3 | 50/50 | 3/3 | 118/118 | 100.0 | 16:54:23 | 0.0 | 0.60 |
| E-5 | 50 | 3 | 148 | 1967 | 148 | 4/4 | 4/4 | 50/50 | 3/3 | 148/148 | 100.0 | 16:55:35 | 0.0 | 5.26 |
| E-6 | 50 | 3 | 136 | 1824 | 125 | 4/4 | 4/4 | 50/50 | 3/3 | 125/136 | 84.2 | 16:54:13 | 0.0 | 1.03 |
| E-7 | 50 | 3 | 112 | 1538 | 112 | 4/4 | 3/4 | 50/50 | 3/3 | 112/112 | 100.0 | 16:54:04 | 0.0 | 0.91 |
| E-8 | 50 | 3 | 109 | 1386 | 109 | 4/4 | 4/4 | 50/50 | 3/3 | 109/109 | 100.0 | 16:53:39 | 0.0 | 8.23 |
| E-9 | 50 | 3 | 141 | 2060 | 141 | 4/4 | 2/3 | 50/50 | 3/3 | 141/141 | 100.0 | 16:54:36 | 0.0 | 4.76 |
| E-10 | 50 | 3 | 130 | 1626 | 130 | 4/4 | 3/4 | 50/50 | 3/3 | 130/130 | 100.0 | 16:55:20 | 0.0 | 2.81 |
| F-1 | - | 5 | 140 | 1451 | 122 | 4/4 | 2/3 | 5/5 | - | 115/140 | 84.4 | 16:55:54 | 2.3 | TL |
| F-2 | - | 5 | 128 | 1993 | 128 | 4/4 | 4/4 | 5/5 | - | 128/128 | 100.0 | 16:54:37 | 0.0 | 1.13 |
| F-3 | - | 5 | 157 | 2049 | 157 | 4/4 | 4/4 | 5/5 | - | 157/157 | 100.0 | 16:55:37 | 0.0 | 33.53 |
| F-4 | - | 5 | 169 | 1601 | 135 | 4/4 | 4/4 | 5/5 | - | 128/169 | 82.5 | 16:55:56 | 1.5 | TL |
| F-5 | - | 5 | 144 | 1658 | 136 | 4/4 | 3/3 | 5/5 | - | 133/144 | 91.7 | 16:54:37 | 2.4 | TL |
| F-6 | - | 5 | 155 | 1580 | 141 | 4/4 | 4/5 | 5/5 | - | 140/155 | 88.9 | 16:53:18 | 0.4 | TL |
| F-7 | - | 5 | 119 | 973 | 92 | 4/4 | 3/4 | 5/5 | - | 92/119 | 58.7 | 16:55:18 | 0.0 | 0.56 |
| F-8 | - | 5 | 113 | 1438 | 113 | 4/4 | 4/4 | 5/5 | - | 113/113 | 100.0 | 16:55:21 | 0.0 | 8.32 |
| F-9 | - | 5 | 130 | 1499 | 117 | 4/4 | 4/4 | 5/5 | - | 114/130 | 92.5 | 16:53:54 | 0.4 | TL |
| F-10 | - | 5 | 114 | 1481 | 114 | 4/4 | 4/4 | 5/5 | - | 112/114 | 99.0 | 16:54:46 | 1.0 | TL |
| G-1 | 50 | 5 | 179 | 2248 | 175 | 4/4 | 4/4 | 5/5 | 44/50 | 163/179 | 92.4 | 16:55:07 | 9.2 | TL |
| G-2 | 50 | 5 | 190 | 2281 | 188 | 4/4 | 4/4 | 5/5 | 49/50 | 184/190 | 92.0 | 16:55:00 | 7.7 | TL |
| G-3 | 50 | 5 | 200 | 1958 | 170 | 4/4 | 4/4 | 5/5 | 50/50 | 170/200 | 61.8 | 16:55:23 | 0.0 | 7.43 |
| G-4 | 50 | 5 | 169 | 1818 | 159 | 4/4 | 4/4 | 5/5 | 50/50 | 159/169 | 83.5 | 16:55:02 | 0.0 | 34.06 |
| G-5 | 50 | 5 | 176 | 2073 | 163 | 4/4 | 3/4 | 5/5 | 50/50 | 157/176 | 71.2 | 16:54:35 | 9.7 | TL |
| G-6 | 50 | 5 | 201 | 2596 | 201 | 4/4 | 3/4 | 5/5 | 40/50 | 188/201 | 98.6 | 16:55:12 | 1.5 | TL |
| G-7 | 50 | 5 | 185 | 2178 | 167 | 4/4 | 4/4 | 5/5 | 50/50 | 167/185 | 61.6 | 16:55:22 | 0.0 | 8.45 |
| G-8 | 50 | 5 | 192 | 1825 | 160 | 4/4 | 4/4 | 5/5 | 48/50 | 144/192 | 80.6 | 16:55:56 | 4.8 | TL |
| G-9 | 50 | 5 | 186 | 2228 | 165 | 4/4 | 3/5 | 5/5 | 50/50 | 164/186 | 77.3 | 16:54:11 | 0.4 | TL |
| G-10 | 50 | 5 | 187 | 2220 | 181 | 4/4 | 4/4 | 5/5 | 48/50 | 176/187 | 97.2 | 16:55:04 | 0.3 | TL |

**CtSpot** - number of spot targets; **CtPoly** - number of polygonal targets; **CtStr** - number of strips; **CtVtw** - number of VTW; **CtStrVtw** - number of strips with at least one VTW within the scheduling period; **CtAcSat** - number of active satellites (satellites with at least one scheduled observation); **CtAcOrb** - number of active orbits (orbits containing scheduled observations / total orbits with imaging opportunities); **CtObSpot** - number of scheduled observations for spot targets; **CtObPoly** - number of scheduled observations for polygonal targets; **CtObStr** - number of scheduled observations for strips; **TotProf** - total obtained profit; **TL** - time limit (set to 1 hour).

Table 3: Comparative results.

| Reference | Instance analysis | | | | Results | | | | | | | | |
|---|---|---|---|---|---|---|---|---|---|---|---|---|---|
| | Ins. | CtPoly | CtStr | Area (km$^2$) | CtAcSat | CtAcOrb | CtObPoly | CtObStr | TotProf(%) | Makespan | Gap(%) | Time(s) |
| (Lu et al., 2023) | Objective function: bi-objective (maximize total observation coverage revenue and minimize makespan) | | | | | | | | | | | | |
| | 1 | 1 | 14 | 31760 | 4/4 | 1/1 | 1/1 | 14/14 | 100.0 | ? | ? | 4778.3 |
| | 2 | 1 | 30 | 201678 | 4/4 | 1/1 | 1/1 | 30/30 | 100.0 | ? | ? | 13044 |
| | 3 | 1 | 46 | 328476 | 4/4 | 1/1 | 1/1 | 46/46 | 100.0 | ? | ? | 43017.4 |
| Ours | Objective function: single-objective (maximize total observation coverage revenue) | | | | | | | | | | | | |
| | 1 | 1 | 22 | 70058 | 3/4 | 1/1 | 1/1 | 22/22 | 100.0 | 07:21:43 | 0.0 | 0.01 |
| | 2 | 1 | 23 | 71451 | 4/4 | 1/1 | 1/1 | 23/23 | 100.0 | 07:22:51 | 0.0 | 0.01 |
| | 3 | 1 | 20 | 69316 | 2/4 | 1/1 | 1/1 | 20/20 | 100.0 | 07:21:04 | 0.0 | 0.01 |
| | 4 | 1 | 21 | 52734 | 4/4 | 1/1 | 1/1 | 21/21 | 100.0 | 07:22:34 | 0.0 | 0.01 |
| | 5 | 1 | 19 | 73114 | 2/4 | 1/1 | 1/1 | 19/19 | 100.0 | 07:20:25 | 0.0 | 0.01 |
| | 6 | 1 | 28 | 76115 | 3/4 | 1/1 | 1/1 | 28/28 | 100.0 | 07:21:49 | 0.0 | 0.01 |
| | 7 | 1 | 17 | 52651 | 3/4 | 1/1 | 1/1 | 17/17 | 100.0 | 07:19:46 | 0.0 | 0.01 |
| | 8 | 1 | 33 | 131070 | 3/4 | 1/1 | 1/1 | 33/33 | 100.0 | 07:20:52 | 0.0 | 0.01 |
| | 9 | 1 | 43 | 260748 | 4/4 | 1/1 | 1/1 | 43/43 | 100.0 | 07:22:16 | 0.0 | 0.71 |
| | 10 | 1 | 34 | 226100 | 4/4 | 1/1 | 1/1 | 34/34 | 100.0 | 07:22:11 | 0.0 | 0.01 |
| | Objective function: lexicographic (maximize total observation coverage revenue and minimize makespan) | | | | | | | | | | | | |
| | 1 | 1 | 22 | 70058 | 2/4 | 1/1 | 1/1 | 22/22 | 100.0 | 07:20:46 | (0.0, 0.4) | TL |
| | 2 | 1 | 23 | 71451 | 3/4 | 1/1 | 1/1 | 23/23 | 100.0 | 07:21:37 | (0.0, 0.7) | TL |
| | 3 | 1 | 20 | 69316 | 2/4 | 1/1 | 1/1 | 20/20 | 100.0 | 07:20:18 | (0.0, 0.4) | TL |
| | 4 | 1 | 21 | 52734 | 3/4 | 1/1 | 1/1 | 21/21 | 100.0 | 07:21:00 | (0.0, 0.6) | TL |
| | 5 | 1 | 19 | 73114 | 2/4 | 1/1 | 1/1 | 19/19 | 100.0 | 07:19:42 | (0.0, 0.4) | TL |
| | 6 | 1 | 28 | 76115 | 3/4 | 1/1 | 1/1 | 28/28 | 100.0 | 07:21:21 | (0.0, 0.1) | TL |
| | 7 | 1 | 17 | 52651 | 2/4 | 1/1 | 1/1 | 17/17 | 100.0 | 07:19:23 | (0.0, 0.5) | TL |
| | 8 | 1 | 33 | 131070 | 3/4 | 1/1 | 1/1 | 33/33 | 100.0 | 07:20:02 | (0.0, 0.6) | TL |
| | 9 | 1 | 43 | 260748 | 4/4 | 1/1 | 1/1 | 43/43 | 100.0 | 07:21:52 | (0.0, 0.9) | TL |
| | 10 | 1 | 34 | 226100 | 4/4 | 1/1 | 1/1 | 34/34 | 100.0 | 07:21:52 | (0.0, 0.9) | TL |

**CtPoly** - number of polygonal targets; **CtStr** - number of strips; **CtAcSat** - number of active satellites; **CtAcOrb** - number of active orbits; **CtObPoly** - number of scheduled observations for polygonal targets; **TotProf** - total obtained profit; **TL** - time limit (set to 10 seconds).

Our multi-objective formulation is defined as follows.

$$\text{Minimize}_{\text{Lex}} \ (-\text{Equation (4)}, C_{\max}) \quad (8)$$

Subject to

$$C_{\max} \geq endOf(S_i), \forall i \in I \quad (9)$$

Equation (5)

Equation (6)

In this experiment, we evaluate the model considering both objective function formulations, the single-objective and the lexicographic, and compare our results with those reported in (Lu et al., 2023). Table 3 summarizes the characteristics of the considered test instances, including both our generated instances and those described in (Lu et al., 2023), and the corresponding computations results.

For the single-objective formulation (Equation (4)), our approach consistently finds the optimal solution in very short computational times. Unlike the results reported in (Lu et al., 2023), we observe that not all satellites are necessarily required to observe a single target, even though each satellite has feasible VTW for all strips. This demonstrates the model's ability to identify efficient observation schedules that avoid redundant satellite usage. Regarding the lexicographic formulation, the solver could not reach optimal solutions within reasonable computational time. However, when imposing a short run time limit of only 10 seconds, the model still produced feasible solutions with optimality gaps below 1%. This indicates that, even though optimality is not achieved, the proposed model can generate high-quality schedules within a very short time frame. Moreover, this run time is significantly lower than the one reported in (Lu et al., 2023), while still allowing an explicit quantification of the solution quality. This performance suggests that our model has the potential to be more computationally efficient.

Overall, these preliminary results show that our exact model has strong potential to compete computationally with non-exact approaches. It not only achieves better computational performance for the tested instances but also provides a direct measure of solution quality, which heuristic or metaheuristic methods typically lack.

# 6 CONCLUSIONS AND FUTURE RESEARCH

This paper addresses the imaging scheduling problem for super-agile Earth observation satellites, where the optimal allocation of strips to satellites within their corresponding observation time windows aims to maximize the total observation profit. Only a few studies in the literature consider the dynamic imaging capabilities of SAEOS and the sequence-dependent transition times between observations, and, to the best of our knowledge, no exact approach had been previously proposed.

The experimental results show that the number of strips to be observed strongly affects the number of visible time windows and, consequently, the computational complexity of the problem. The proposed model proved to be highly efficient for small-sized instances, achieving optimal solutions within short computational times. Moreover, the comparative results indicate that the proposed approach attains higher computational performance than current state-of-the-art non-exact methods, while explicitly providing an evaluation of solution quality. For larger instances, the model is not always able to reach proven optimality within reasonable time limits. However, near-optimal solutions can be obtained within just a few seconds.

This work represents the first contribution in the literature that applies CP to the SAEOS-ISP, and the first exact approach proposed for this problem. As future work, we intend to develop new methodologies to address the scalability of the model and to extend the formulation to incorporate additional challenges inherent to Earth observation satellite scheduling, namely energy and onboard memory constraints.

# ACKNOWLEDGEMENTS

This research is sponsored by national funds through FCT – Fundação para a Ciência e a Tecnologia, under projects UID/00285/2025 and LA/P/0112/2020, and by the Operational Programme for Competitivity and Internationalization of Portugal 2020 Partnership Agreement (PRODUTECH4S&C), grant number POCI-01-0247-FEDER-046102. All supports are gratefully acknowledged. No generative artificial intelligence tools were used in the preparation of this paper. This work has been accepted for publication in the Proceedings of the International Conference on Operations Research and Enterprise Systems (ICORES 2026).